\documentclass[prl,twocolumn,
superscriptaddress,showpacs,floatfix]{revtex4-1}
\usepackage{hyperref}
\usepackage[british]{babel}
\usepackage[utf8]{inputenc}
\usepackage{amsmath}
\usepackage{amssymb}
\usepackage{mathtools}
\usepackage{graphicx}
\usepackage{xcolor}

\DeclareMathAlphabet{\mathpzc}{OT1}{pzc}{m}{it}

\begin{document}

\title{From black hole spectral instability to stable observables}
\author{Th\'eo Torres}
\email{torres@lma.cnrs-mrs.fr}
\affiliation{Aix Marseille Univ., CNRS, Centrale Marseille, LMA UMR 7031, Marseille, France.}
\date{\today}

\begin{abstract}
The quasi-normal mode (QNM) spectrum of black holes is unstable under small perturbation of the potential and has observational consequences in time signals. 
Such signals might be experimentally difficult to observe and probing this instability will be a technical challenge. 
Here we investigate the spectral instability of time-independent data. 
This leads us to study the Regge Poles (RP), the counterparts to the QNMs in the complex angular momentum plane. We present evidence that the RP spectrum is unstable but that not all overtones are affected equally by this instability.  
In addition, we reveal that behind this spectral instability lies an underlying structure. The RP spectrum is perturbed in such a way that one can still recover \emph{stable} scattering quantities using the complex angular momentum approach. 
Overall, the study proposes a novel and complementary approach on the black hole spectral instability phenomena which allows us to reveal a surprising and unexpected mechanism at play which protects scattering quantities from the instability.
\end{abstract}

\maketitle\noindent

\textbf{Introduction}.--- 
Resonant phenomena are found ubiquitously in various branches of physics, from mechanical vibrations and electromagnetism to quantum mechanics and gravitational interactions. Resonances constitutes a powerful mathematical framework that reveal striking similarities and connections across fields.

In non-conservative systems, described in the general framework of non-Hermitian physics~\cite{ashida2020non}, the spectrum of resonance may undergo the elusive phenomena of spectral instability. Spectral instability refers to the susceptibility of a system's spectrum to perturbations. In other words, a small modification to the system's environment might change drastically the spectrum of its resonances.

Within this context, it is important to note that resonances provides a tool to describe physical phenomena but they are not the observables measured in experiments. A
close consideration of the link between spectral instabilities and physical observables is therefore warranted and it is the main purpose of this Letter.

We will address this issue by focusing on a specific system: a classical Schwarzschild black hole (BH).

Motivated by early works by Nollert and Price~\cite{doi:10.1063/1.532698} and others~\cite{AGUIRREGABIRIA1996251,Vishveshwara:1996jgz}, Jaramillo \textit{et al.} have recently identified and described the quasi-normal mode spectral instability of BHs~\cite{Jaramillo:2020tuu}. 
Quasi-normal modes (QNMs) are a set of damped sinusoids characterised by discrete complex frequencies: the QNM spectrum. 
QNMs are of prime importance in BH physics as they relate to the late-time GW signal, sometimes refered to as the ``ringdown'', emitted after BH mergers~\cite{Berti:2009kk,Konoplya:2011qq,PhysRevD.105.064046}.

In~\cite{Jaramillo:2020tuu,alsheikh:tel-04116011,Gasperin:2021kfv}, it was shown that the QNMs spectrum of BHs exhibits an instability and may deviate drastically if a small perturbation is added to the BH environment. 
The natural question of the observational consequences of this instability was quickly asked and signs of it were found in the natural physical quantity associated to QNMs: the ringdown of perturbed BH~\cite{Jaramillo:2021tmt}.
Soon after, it was shown that some perturbations of the BH surroundings could also destabilise the fundamental QNMs~\cite{Cheung:2021bol}, which can also be seen in ringdown signals~\cite{Berti:2022xfj}.

Naturally, the recent discovery that the QNM spectrum of BHs is unstable under certain small perturbations of the potential poses a challenge to the influential \emph{black hole spectroscopy} proposal~\cite{Dreyer:2003bv,Berti:2005ys,Berti:2016lat,Baibhav:2023clw}: that the detection of multiple quasinormal modes can be used to test the no-hair theorem~\cite{Gossan:2011ha,Isi:2019aib}.

In this context and in light of the
incredible technological progresses, exemplified by the first picture of a BH shadow by the Event Horizon Telescope~\cite{EventHorizonTelescope:2019dse,Vagnozzi:2022moj} and the future LISA mission~\cite{LISA},
modelling and understanding wave propagation around BHs in our inhomogeneous and matter-filled Universe is therefore of the utmost importance.


While considerable progress has been made regarding the implication of a BH spectral instability on time-dependent data, no attention has been paid to the effect of this instability on \emph{time-independent} observables. 
Such time-independent quantities (deflection angle, scattering and absorption cross section) are known to be linked to the spectrum of resonances~\cite{Decanini:2002ha,Stefanov:2010xz,PhysRevD.81.104039,Decanini:2011xi}. 

A natural framework to address the link between spectral instability and scattering quantities is that of the \emph{Regge Poles} (RPs). 
RPs are the counterpart to QNMs and they are a different point of view of the BH resonances. 
QNMs are resonances with a real angular momentum and complex frequency while RPs are resonances with a real frequency and a complex angular momentum.
RPs provide a direct link to scattering and absorption via the complex angular momentum (CAM) approach, which allows one to reconstruct time-independent quantities from the RPs spectrum~\cite{Decanini:2011xi,Decanini:2011xw,Folacci:2018sef}.
Such approach has been successfully applied to BHs and compact objects in various contexts~\cite{Andersson_1994,Decanini:2002ha,Folacci:2018sef,Folacci:2019cmc,OuldElHadj:2019kji}.

Considering the strong link between QNMs and RPs~\cite{Decanini:2002ha}, it is natural to ask the following questions: is the RP spectrum unstable? If so, what are the observational consequences on quantities constructed from them?

In this Lettter, we link features and properties of time-independent scattering quantities to the spectral instabilities of BHs. We present evidences of the following observations: i) the RP spectrum is unstable, and ii) the spectral instability does not manifest itself in the observable scattering quantities linked to the RPs and one can reconstruct them using either the perturbed or unperturbed RP spectrum.

The observation of the stability of physical observables despite the instability of an underlying spectrum is particularly surprising as it implies the existence of compensating mechanisms and suggests that the RP spectrum is not modified in an arbitrary manner.


We will work in geometrical units where $G=c=1$.
\newline

\textbf{Model and Methods}.---
Efforts in studying wave propagation in BH surrounded by an environment has already been made, and authors have considered wave scattering by BH surrounded by matter with specific forms~\cite{Barausse:2014tra,Barausse:2014pra}. Here, we adopt a more general approach and consider a BH surrounded by a generic perturbation.
We consider a scalar field propagating over a Schwarzschild background. Upon separation of variables, the radial profile of the scalar field, $\phi(r)$, satisfies the Regge-Wheeler equation~\cite{Chandrasekhar:1985kt}:
\begin{equation}\label{eq:RW_eq}
    \frac{d^2\phi}{dr_*^2} + (\omega^2 - V_\ell)\phi = 0. 
\end{equation}
In the above, $V_\ell$ is the effective potential for the $\ell$ mode given by
$V_\ell = V_\ell^0 + \delta V$,
with 
\begin{equation}
    V_\ell^0 = \left( 1 - \frac{2M}{r}\right)\left( \frac{\ell(\ell + 1)}{r^2} + \frac{2M}{r^3}\right),
\end{equation}
and $\delta V$ is a perturbation of the radial potential.
Here we are interested in the way that the instability that afflicts the QNM spectrum may also manifest in the RP spectrum, we choose a Gaussian perturbation $\delta V$ which causes a drastic modification of the QNMs spectrum when placed far from the scattering centre. We follow~\cite{Cheung:2021bol} and choose
\begin{equation}\label{eq:perturbation}
    \delta V = \begin{cases}
    \epsilon \exp\left(-\frac{(r_* - a)^2}{\sigma^2}\right) \text{ if } r \leq a + w,\\
    \qquad  \quad 0 \qquad \qquad \text{ if } a + w < r.
    \end{cases}
\end{equation}

We note that this model is an \textit{ad hoc} perturbation. While it may relate to some matter surrounding an isolated black hole~\cite{Barausse:2014tra},
it does not correspond to a solution to the Einstein’s equations modeling an astrophysical BH surrounded by matter. 
In addition, while the amplitude of $\delta V$ is very small compared to potential barrier, and can be considered as a small perturbation, it may lead to a large energy content for large values of $a$~\footnote{The consideration of various norms and the evaluation of the pseudo-spectrum associated to operator linked to the RP is underway and will be presented in a future work.}.
The motivation for the adopted model is qualitative and the goal is to investigate the link between spectral instabilities and physical observables.
The cut is introduced to calculate high overtones of the RP spectrum efficiently with the method described below. For \mbox{$w \gg \sigma$}, the cut does not qualitatively affect the RP spectrum, as shown in the Supplemental Material.

Resonances are defined as solutions to Eq.~\eqref{eq:RW_eq} which satisfy purely outgoing boundary conditions at infinity and ingoing at the horizon, corresponding to energy leaving the system. 
Practically, the spectrum of resonances is determined from the condition $W(\phi^{in}, \phi^{up}) = 0$, where $W$ is the Wronskian of the solutions $\phi^{in}$ and $\phi^{up}$ which separately satisfy the boundary condition at infinity and at the horizon respectively. The spectrum is a discrete set labeled by the integer $n$ called the overtone number. For real $\ell$, the spectrum is the set $\omega_{n} \in \mathbb{C}$ and correspond to QNMs while for real $\omega$, the spectrum lies in the complex $\ell$-plane and correspond to the RPs.

In order to compute the RP spectrum of the perturbed BH, we apply an extension of the well known continued fraction method~\cite{doi:10.1098/rspa.1985.0119}. 
This method consists of constructing a series solution at an arbitrary point $b$, rather than at the horizon, and to construct a continued fraction from the value of the field and its derivative at this point, which are obtained by solving numerically Eq.~\eqref{eq:RW_eq} subject to ingoing boundary condition at the horizon. 
This method was originally developed to study resonances of compact objects and allows one to calculate high overtones of the resonance spectrum while modifying at will the potential for $r<b$~\cite{Benhar:1998au,OuldElHadj:2019kji,Torres:2022fyf} (see Supplemental Material). We have checked that this method and a direct integration scheme~\cite{Chandrasekhar:1975zza} give the same spectrum (see Supplemental Material).
\newline

\textbf{Results}.--- We now consider a particular perturbation of the radial potential, fixing $2M=1$ and given by Eq.\eqref{eq:perturbation} with the parameters: $\epsilon = 10^{-5}$, $a=10$, $\sigma = 0.1$ and $w = 20\sigma$. With this set of parameters, the fundamental QNM frequency is destabilised, as expected from~\cite{Cheung:2021bol} (see Supplemental material). We now investigate if this QNM spectral instability translates into the complex angular plane. Our results can be cast into two main statements:

\emph{i) Instability of the RP spectrum:} We begin by computing the RP spectrum of the perturbed BH, which is displayed in red in Fig.~\ref{fig:RP_spectrum} and given in Table.~\ref{tab:RP_spectrum} in the Supplemental Material. 
The spectrum and in particular the high overtones is clearly unstable and deviates from the unperturbed spectrum shown in blue. 
We can identify three branches in the perturbed spectrum: 1) an ``original'' branch made of low overtones which follow the unperturbed spectrum (represented in red squares) 2) an ``inner'' branch (red diamonds) with lower real parts and regularly spaced imaginary part and 3) an ``outer'' branch (red triangles) with larger real parts and an increasing spacing of imaginary parts. Some properties of the RP spectral instability are investigated in the Supplemental material and in a future work~\cite{Dolan&Torres}. 
Of particular interest is the observation that, for the perturbation considered, the fundamental RP and the first few overtones remain stable while the fundamental QNM is already destabilised.

\begin{figure}
    \centering
\includegraphics[scale=0.75,trim=0.7cm 0 0 0]{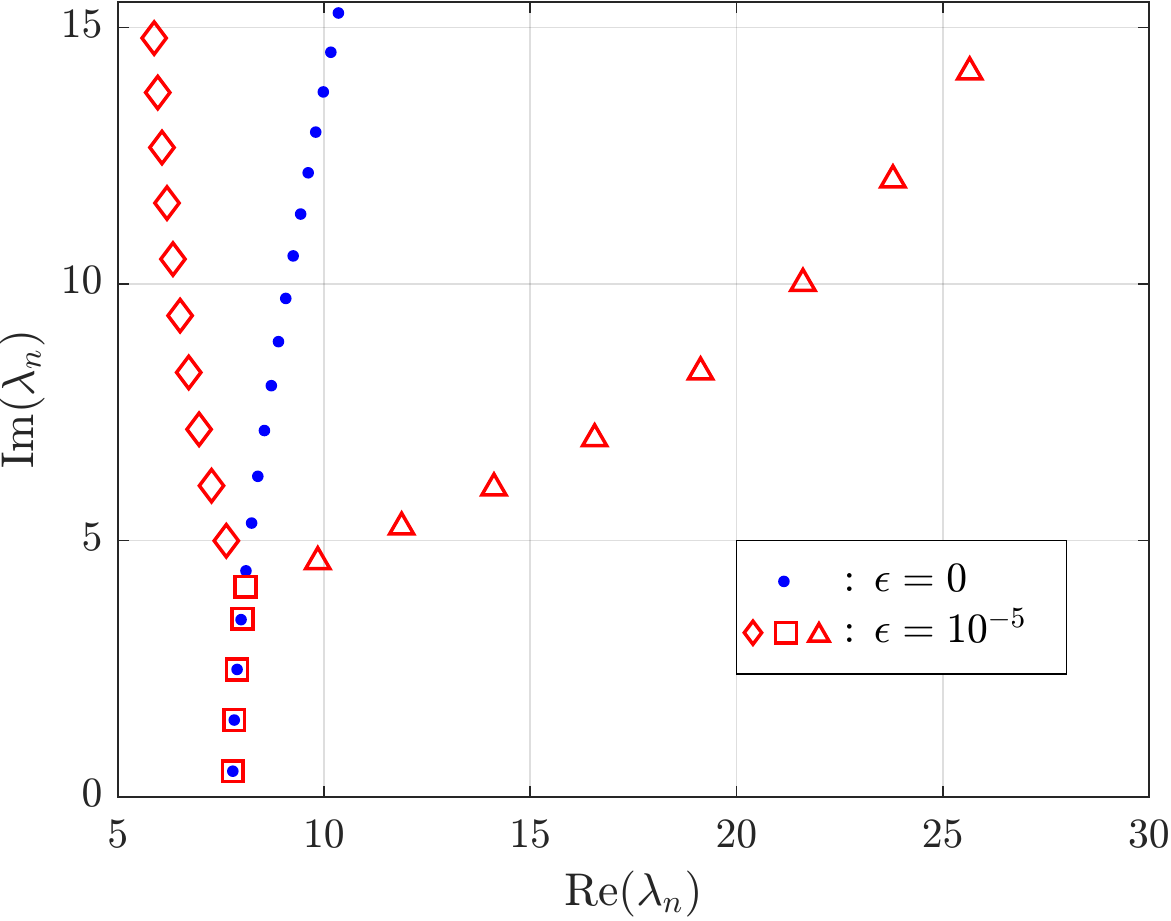}
    \caption{RP spectrum of the perturbed BH (red squares, diamonds and triangles) and of the unperturbed BH (blue dots) for the frequency $2M\omega = 3$. We can distinguish 3 branches in the perturbed spectrum, represented by the different markers.The red squares follow the unperturbed spectrum while the diamonds and the triangles describe the inner and outer branches respectively. The parameters used in $\delta V$ are $\sigma = 1/10$, $w=20\sigma$ and $a=10$.}
    \label{fig:RP_spectrum}
\end{figure}

\emph{ii) Stability of physical observables:} We have established that the RP spectrum is unstable and answered the first question raised at the beginning of this Letter. 
We now turn our attention to the second question and address whether or not the spectral instability manifests itself on observable quantities linked to the RP spectrum.

There are two fundamental scattering quantities that can be related to the RP spectrum: 1) the scattering cross section, $\frac{d\sigma}{d\Omega}$, and 2) the absorption cross section, $\sigma$~\footnote{Note that we have used the standard notation for the scattering and absorption cross section but the symbol should not be mixed up and the scattering cross section is not the derivative per solid angle of the absorption cross section}.
The scattering cross section at a fixed frequency is constructed from the scattering amplitude $f(\omega,\theta)$ via $\frac{d\sigma}{d\Omega} = |f(\omega,\theta)|^2$.
Using the partial wave method, the scattering amplitude and the absorption cross section can be computed from a summation over the different $\ell$ components and their ingoing/outgoing amplitudes at infinity. ~\cite{doi:10.1063/1.1664470,pike2001scattering}

The scattering amplitude and the absorption cross section can be related to the RP spectrum via the complex angular momentum (CAM) approach~\cite{Andersson_1994,Folacci:2019cmc,Decanini:2011xi}.
The main idea of the CAM approach is to turn the sum over $\ell$ modes into an integral over a continuous variable, $\lambda = \ell +1/2$, in the complex angular momentum plane.
Via the CAM approach, the scattering quantities can be exactly expressed as
\begin{equation}\label{eq:CAM}
    f(\omega,\theta) = f^{B} + f^{RP}, \text{ and } \sigma = \sigma^{B} + \sigma^{RP}
\end{equation}
where the superscript $B$ is a contribution from a background integral and the superscript $RP$ denotes the contribution from the RPs. The background contribution is dominant at low frequency while it is the RP contribution that dominates at high-frequencies, independently of the perturbation of the potential.

Eq~\eqref{eq:CAM} makes a direct link between the RP spectrum and physical observables; and it is this link that we should investigate now.

\begin{figure}
    \centering
    \includegraphics[scale=0.65,trim=0.7cm 0 0 0]{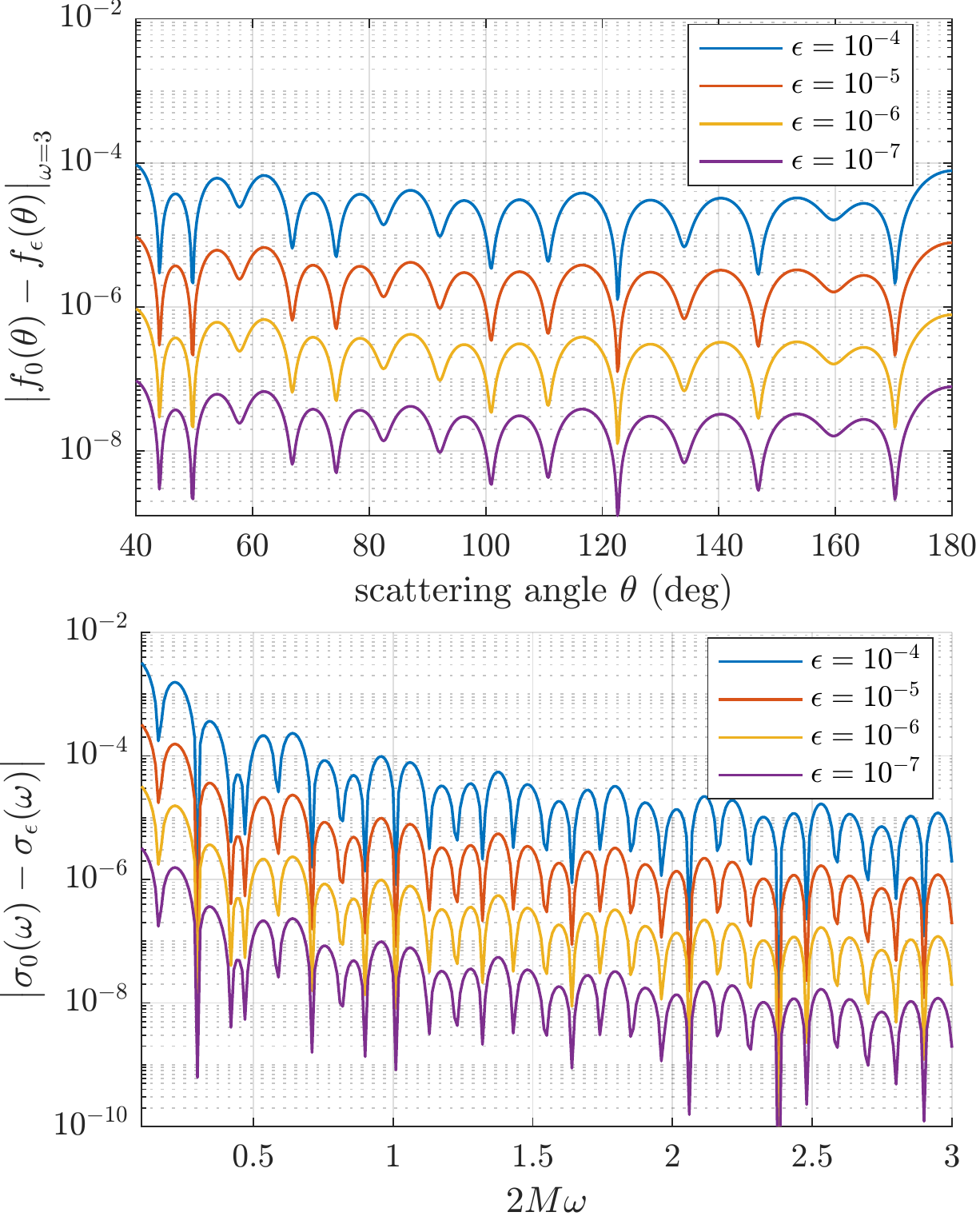}
    \caption{Difference between the unperturbed and perturbed physical observable. The top panel depicts the difference in the scattering amplitude for the frequency $2M\omega = 3$ while the bottom panel represents the difference in the absorption cross section.}
    \label{fig:diff}
\end{figure}

First, we observe that the scattering quantities are stable under the perturbation considered here. If we denote $f_{\epsilon}$ ($\sigma_\epsilon$) the scattering amplitude (absorption cross section) of the perturbed potential and $f_{0}$ ($\sigma_0$) the one of the unperturbed potential, one can see that the difference between the two is of order $\epsilon$. This is represented in Fig.~\ref{fig:diff} which depicts the two differences computed numerically using the partial wave method in conjunction with the reduction series technique~\cite{Dolan:2008kf,Dolan:2017rtj}.
Note also that the difference scales linearly with epsilon, indicating that non-linear contributions in $\epsilon$ are negligible. 

\begin{figure}
    \centering
    \includegraphics[scale=0.82,trim=0.5cm 0 0 0]{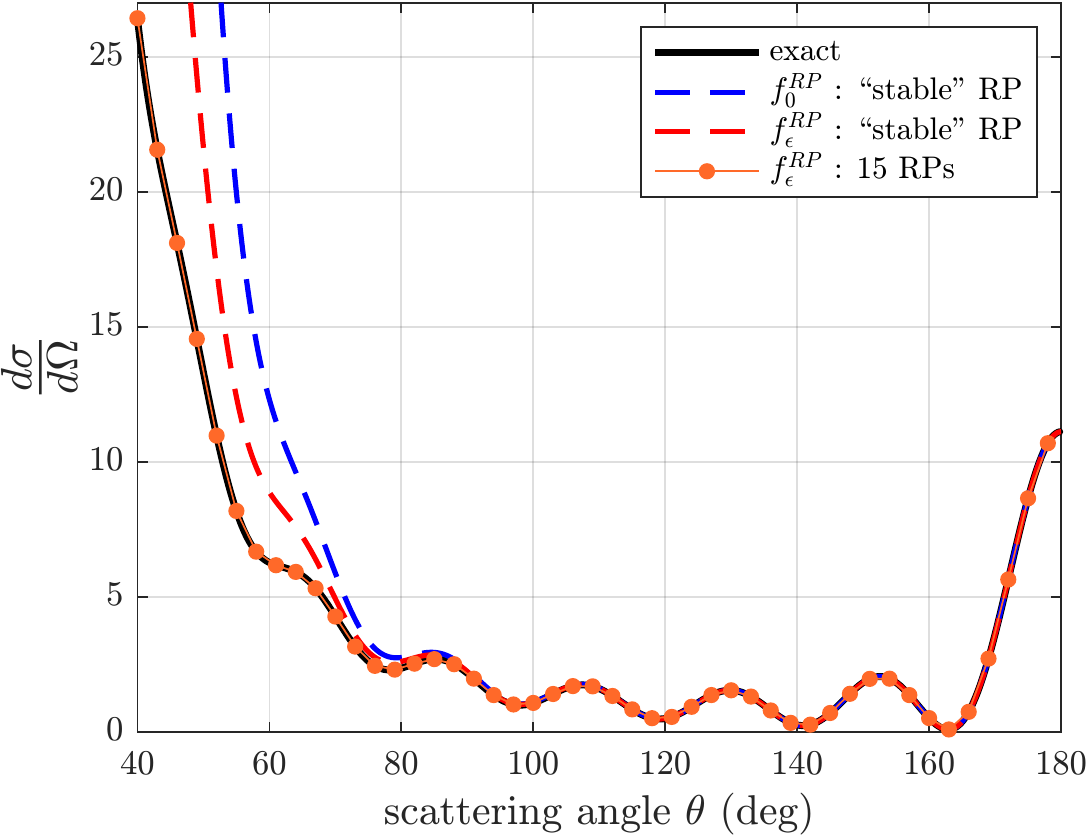}
    \caption{Scattering cross section and its reconstruction from the RP spectrum. The solid black line represents the scattering cross section of the unperturbed BH computed from the partial wave method. The dashed curves are the RP reconstruction where only the first 5 overtones of the unperturbed (in blue) and perturbed (red) spectra, corresponding to the original branch are included. The orange curve is the reconstruction where we have included the first 15 perturbed RPs. Note the remarkable agreement between the unperturbed scattering cross section and the reconstruction from the unstable spectrum.}
    \label{fig:scat_cross_RP}
\end{figure}

A key result of this Letter is that one can reconstruct \textit{stable} physical quantities from either the unperturbed or the perturbed RP spectrum. 
This is best illustrated in Fig.~\ref{fig:scat_cross_RP}, where the unperturbed scattering cross section is reconstructed (dotted orange line) from the first $15$ overtones of the perturbed RP spectrum  presented in Fig.~\ref{fig:RP_spectrum} (the reconstruction from the stable spectrum can be found in~\cite{Folacci:2019cmc}). 
We note that the contribution of the background integral is negligible for the frequency considered here and
the RP contribution is calculated directly from the RP spectrum and its associated residues, $r_n(\omega)$~\cite{Folacci:2019cmc}
\begin{equation}\label{eq:fRP}
    f^{RP}(\omega,\theta) = -\frac{i\pi}{\omega}\sum_{n=0}^{\infty} \frac{\lambda_n(\omega) r_n(\omega)}{\cos(\pi \lambda_n(\omega))}\times P_{\lambda_n(\omega)-1/2}(-\cos(\theta)).
\end{equation}

As we pointed out above, the RP spectrum presents a relatively ``stable'' part where the perturbed and unperturbed RP spectrum coincide on the low overtones.
One might expect those low overtones to dominate the reconstruction of the scattering cross section, much like the fundamental QNMs dominates the ringdown signal.
This is however not the case, and while the low overtones account for the large angular behaviour, one does need to include the full perturbed RP spectrum to reconstruct the scattering cross section.
This is illustrated in Fig.~\ref{fig:scat_cross_RP}, where the RP contribution from the ``stable'' part of the spectrum is isolated (shown in dashed lines). 
We underline again here the remarkable agreement between the scattering cross section of the unperturbed BH and the reconstruction from the drastically different unstable RP spectrum. 

\begin{figure}
    \centering
    \includegraphics[scale=0.55,trim=0.5cm 0 0 0]{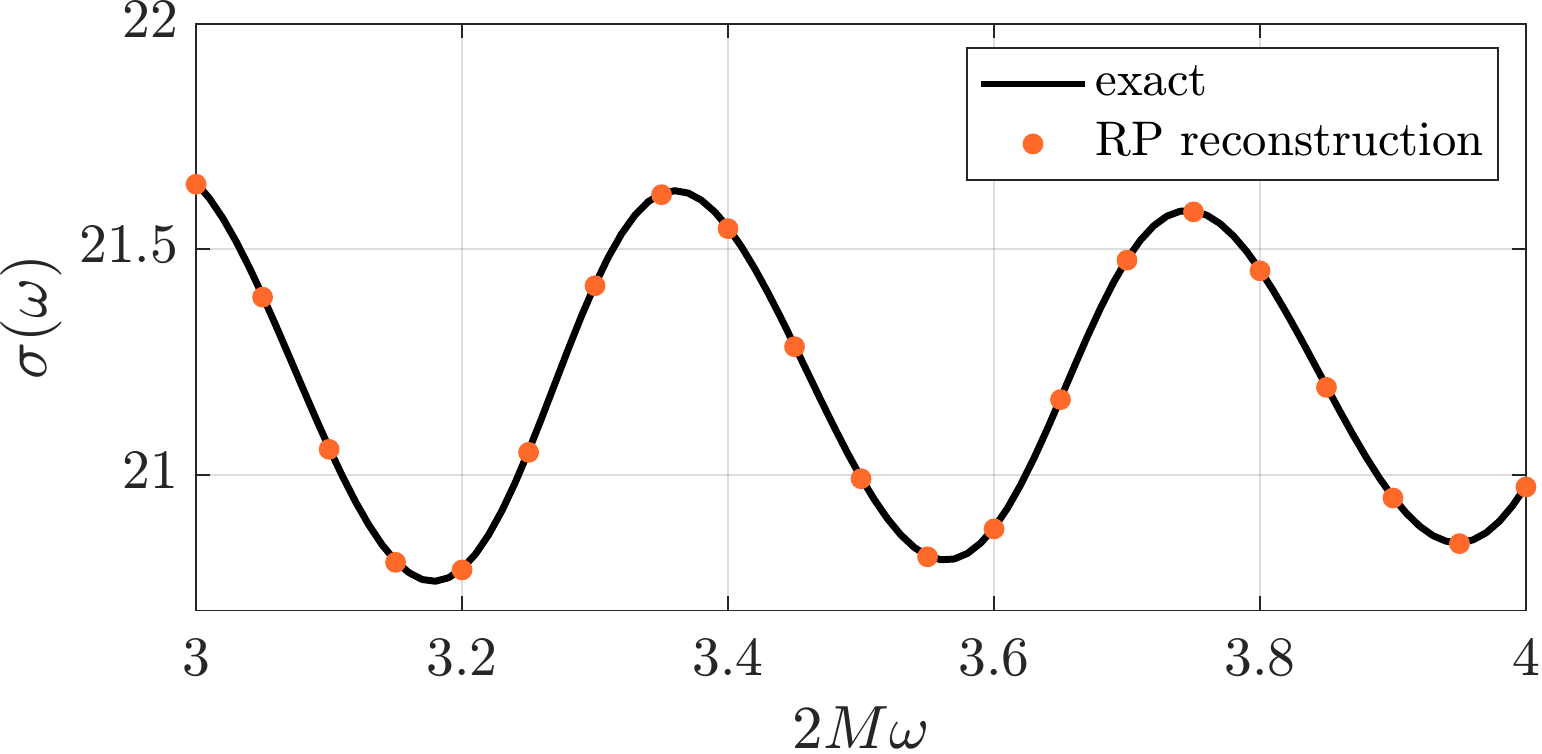}
    \caption{Absorption cross section and its reconstruction from the RP spectrum. The solid black line represents the absorption cross section of the unperturbed BH computed from the partial wave method. The dotted curve is the RP reconstruction including the first 8 perturbed RPs.}
    \label{fig:abs_cross_RP}
\end{figure}

A similar reconstruction of the stable absorption cross section is presentend in Fig.~\ref{fig:abs_cross_RP}. 
In the high frequency regime, the background contribution can be approximated by $\sigma^B \sim 27 \pi M^2 $ and therefore we have that \mbox{$
\sigma \sim 27\pi M^2 + \sigma^{RP}$}, with the RP contribution given by
\begin{equation}\label{eq:sigma_CAM}
    \sigma^{RP} = -\frac{4\pi^2}{\omega^2}\textrm{Re}\left( \sum_{n=0}^{\infty} \frac{\lambda_n(\omega) \gamma_n(\omega)e^{i \pi [\lambda_n(\omega) - 1/2]}}{\sin(\pi(\lambda_n(\omega)-1/2))}
    \right),
\end{equation}
where $\gamma_n$ are the residues of the gray-body factor $\Gamma_\ell(\omega)$~\cite{Decanini:2011xi}.
\newline

\textbf{Discussion}.---
Motivated by understanding the physical consequences of the BH spectral instability, we have adopted a different point of view on the problem, that of the RPs and of the CAM method.
This approach allowed us to unveil another facet of the instability by exploring it in the complex angular momentum plane.

We have presented evidences that the RP spectrum of BH is unstable.
This observation reinforces the strong link already established between QNMs and RPs.
Our analysis however, reveals that the link between those two facets of resonances might not be so straightforward and it is necessary to study them separately.

While a detailed study of the properties of the RP spectrum will be presented in coming works, we have pointed out that the RP spectrum exhibits an increased robustness to perturbation as compared to QNMs.
By this, we mean that under situation where the fundamental QNM frequency is strongly destabilised, the fundamental RP remains stable.
A potential explanation for this superior robustness is the direct link that exists between the low overtones of the RP spectrum and physical effects, associated with particle trajectories in a black hole space-time~\cite{PhysRevD.30.295,Cardoso:2008bp,Andersson_1994}

Finally, we have shown that while the RP spectrum is unstable, its impact on physical observable quantities, the scattering cross section and the absorption cross section, is minimal. This is in sharp contrast with the case of time-dependent quantities in which signatures of the QNM spectral instability can clearly be extracted~\cite{Berti:2022xfj,Jaramillo:2021tmt}.
This observation reveals that a surprising and unexpected compensation mechanism is at play in the spectral instability phenomena.
In particular, Eqs.~\ref{eq:fRP} and~\ref{eq:sigma_CAM} are two highly non-trivial combinations of unstable individual spectral quantities, that overall remain stable.
To the best of our knowledge, this is the first evidence of the presence of such structure in BH spectral instabilities.
Finding all invariant spectral combination and their properties is an promising avenue for future research and may provide valuable insight into our understanding of spectral instabilities across fields.

An another interesting application of our results would be to look for signs of structure in the QNM spectral instability. 
It was already noted that the QNM spectral instability will manifest on late time-scale and therefore, early time signals would be stable under perturbation~\cite{Berti:2022xfj,Barausse:2014tra,Jaramillo:2022kuv}.
It would however be remarkable to identify invariant quantities linked to the QNM spectrum. A potential candidate for such quantity could be the deflection angle in the strong deflection limit which is directly linked to the scattering cross section and the QNM spectrum~\cite{Raffaelli:2014ola,Jusufi:2020dhz}

Ultimately, while we have focused on the link between spectral stability and observable in the context of BH physics, we anticipate that our observations may have applications in other fields where the CAM approach is applicable such as atomic and molecular collisions~\cite{FT9908601627} or particle physics~\cite{gribov_2003} where signs of RP spectral instability may have been seen~\cite{Hiscox:2011wpd}.

\section*{Acknowledgments}
I would like to thank Jos\'e-Luis Jaramillo and Sam Dolan for inspiring and insightful discussions on various aspects of the project as well as valuable inputs on the manuscript. I am grateful to Antonin Coutant and Sam Patrick for discussing and commenting early versions of the manuscript. I would also like to thank Bruno Lombard for bringing up to me the notions of eigenmodes and resonant modes. I would like to thank Mohamed Ould El Hadj for introducing me to the subject of Regge Poles and his insights on various numerical aspects.
Finally, I thank two anonymous referees for their comments which have improved the original manuscript.

\bibliography{bibli}

\newpage
\newpage
\section*{Supplemental Material}

\subsection{Numerical method}
In order for the paper to be self-contained, we recall here the numerical method used to compute the RP spectrum.
We follow the method
used in~\cite{OuldElHadj:2019kji,Torres:2022fyf} who calculated the RP spectrum
of scalar and gravitational waves (in the axial sector) for a gravitating compact body and for a scalar field in a dirty black hole spacetime.
The method is an extension of the original continued fraction method developed
originally by Leaver~\cite{Leaver:1985ax,leaver1986solutions}.

The method involves writing the solution to the wave equation
\eqref{eq:RW_eq} as a power series around a point $b$ located
outside of the perturbation (hence the need for the perturbation to have compact support),
\begin{equation}\label{ansatz_CFM}
\phi_{\omega,l}(r) = e^{i\omega r_*(r)} \sum_{n=0}^{+\infty} a_n \left(1 - \frac{b}{r}\right)^n\,,
\end{equation}
where the coefficients $a_n$ obey a four-term recurrence relation:
\begin{equation}\label{Recurrence_4_terms}
\alpha_n a_{n+1} + \beta_n a_{n} +\gamma_{n} a_{n-1} +\delta_{n} a_{n-2}  = 0,
\quad \forall n\geq 2 ,
\end{equation}
where
\begin{subequations}
\begin{eqnarray}\label{Coeffs_3_termes}
&& \alpha_n = n (n+1)\left(\!1-\frac{2M}{b}\!\right),    \\
&& \beta_n  = n\left[\left(\!\frac{6M}{b}-2\!\right)n + 2ib\omega\right] ,  \\
&& \gamma_n = \left(\!1-\frac{6M}{b}\!\right)n(n-1)-\frac{2M}{b}-\ell(\ell+1) ,  \\
&& \delta_n = \left(\!\frac{2M}{b}\!\right)\left(n-1\right)^2 .
\end{eqnarray}
\end{subequations}
The initialisation coefficients, $a_0$ and $a_1$, are found directly from
\eqref{ansatz_CFM},
\begin{eqnarray}\label{Initial_Conds}
&& a_0 = e^{-i\omega r_*(b)}\phi_{\omega,\ell}(b) ,  \\
&& a_1 = b e^{-i\omega r_*(b)}\Bigg{[}\frac{d}{dr}\phi_{\omega,\ell}(r)
-\frac{i\omega b}{b-2M}\phi_{\omega,\ell}\Bigg{]}_{r=b}.
\end{eqnarray}
The coefficients $a_0$ and $a_1$ are found numerically by integrating
\eqref{eq:RW_eq} from the horizon up to $r = b> a+w$.

We then perform a Gaussian elimination step in
order to reduce the 4-term recurrence relation to a 3-term recurrence relation:
\begin{equation}
\hat{\alpha}_n a_{n+1} + \hat{\beta}_n a_n + \hat{\gamma}_n a_{n-1} = 0,
\end{equation}
where we have defined the new coefficients, for $n\geq 2$:
\begin{subequations}
\begin{eqnarray}
&&\hat{\alpha}_n  = \alpha_n, \\
&& \hat{\beta}_n  = \beta_n - \hat{\alpha}_{n-1}\frac{\delta_n}{\hat{\gamma}_{n-1}},
\ \text{and} \\
&& \hat{\gamma}_n = \gamma_n - \hat{\beta}_{n-1}\frac{\delta_n}{\hat{\gamma}_{n-1}}.
\end{eqnarray}
\end{subequations}
The series expansion \eqref{ansatz_CFM} is convergent outside the perturbation provided
that $a_n$ is a minimal solution to the recurrence relation and $b/2<a+w<b$
\cite{Benhar:1998au}.
The existence of a minimal solution implies that the following continued fraction holds:
\begin{equation}\label{eq:CF}
\frac{a_1}{a_0} = \frac{-\hat{\gamma}_1}{\hat{\beta}_1 -}\,
\frac{\hat{\gamma}_2 \hat{\alpha}_1}{\hat{\beta}_2 -} \,
\frac{\hat{\gamma}_3 \hat{\alpha}_2}{\hat{\beta}_3 -} ...
\end{equation}
The above relation (or any of its inversions) is the equation, written in the standard form of continued fractions, we are solving in
order to find the RP spectrum.
In practise, we fix $\omega$, and define a function
$f(\ell\in \mathbb{C})$, which
gives the difference between the left-hand side and right-hand side of the
condition \eqref{eq:CF}. We then find the zeros of the function $f$ starting from an initial guess.

\subsection{Regge Pole Spectrum}

We provide here some more details one the RP spectrum presented in the main text.
In particular we provide data for the first overtones plotted in Fig.~\ref{fig:RP_spectrum} and we then investigate the effect of the cut perturbation on the spectrum.
\subsubsection{Regge Pole data}
Tab.~\ref{tab:RP_spectrum} presents the value of the first 15 overtones of the RP spectrum presented in the main text, corresponding to the frequency $2M\omega=3$ and with the perturbation parameters $\sigma=0.1$, $a=10$ and $\epsilon = 10^{-5}$.
\begingroup
\begin{table}
\caption{\label{tab:RP_spectrum} The lowest RPs $\lambda_{n}(\omega)$ for the
scalar field at frequency $2M\omega=3$ for the perturbed Schwarzschild BH. The RPs are ordered by increasing imaginary parts.}
\smallskip
\centering
\begin{ruledtabular}
\begin{tabular}{ccc}
 $n$  &branch & $\lambda_n$
\\ \hline
0 & original & $7.790119 + 0.500553i$  
\\
1 & original & $7.825561 + 1.497742i$  
\\
2 &original & $7.892958 + 2.485648i$  \\
3 &original & $8.028453 + 3.471837i $  \\
4 &original & $8.095281 + 4.095891i$  \\
5 &outer & $9.849842 + 4.586756i$  \\
6 &inner& $7.632469 + 4.992255i$  \\
7 &outer& $11.882802 + 5.260555i$  \\
8 &outer & $14.121423 + 6.022191i$  \\
9 &inner & $7.273894+ 6.065747i$ \\
10 &outer & $16.561943 + 6.982281i$  \\
11 &inner & $6.972859 + 7.165046i$  \\
12 &inner & $6.721543 + 8.273551i$  \\
13 &outer & $19.129862 + 8.284576i$  \\
14 &inner & $6.512876 + 9.381536$  \\
15 &outer & $21.612042 + 10.010909i$ 
\end{tabular}
\end{ruledtabular}
\end{table}
\endgroup

\subsubsection{Effect of the cut}

In the main text, we have calculated the RP spectrum using an extension of the continued fraction method. 
The method used is particularly useful as it allows one to compute relatively high-overtones while having the freedom to modify at will the potential. 
A limitation of the method is that the perturbation must have compact support which implies that we had to cut the perturbation. 
On the other hand, the direct integration scheme does not have this limitation and allows us to explore the effect of the cut in the perturbation on the RP spectrum.
In addition, the RP being purely oscillatory at infinity and one can use the direct integration method to compute the first few overtones with relatively low computational cost.

\begin{figure}
    \centering
\includegraphics[scale=0.8,trim=1cm 0 0 0]{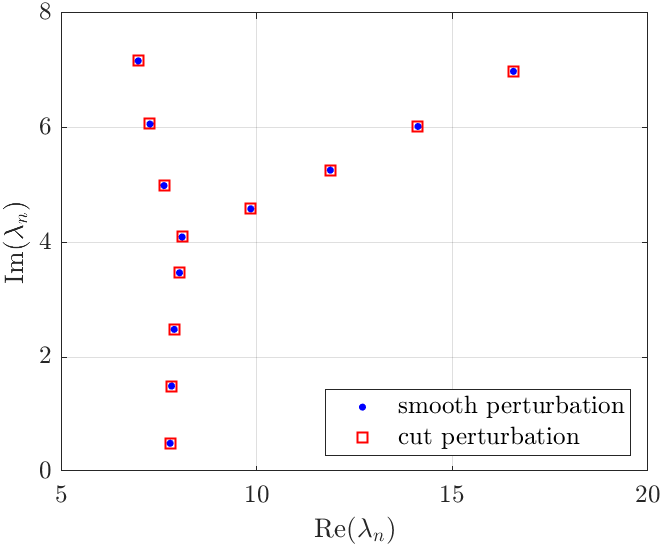}
    \caption{Comparison between the RP spectrum when the Schwarzschild potential is perturbed with a smooth perturbation (in blue dots) and with the cut perturbation used in the main text (in red squares). The two spectrum are remarkably close to each other, showing that the instability is due to the presence of the bump and not of the cut.}
    \label{fig:effect_cut_RP_spectrum}
\end{figure}

Fig.~\ref{fig:effect_cut_RP_spectrum} presents the RP spectrum of the uncut perturbation and compares it with the one of the cut perturbation presented in the main text. 
We see that the two spectra superimposes, implying that the RP spectral instability and its structure is due to the presence of the bump and not of the cut.

We note that the continued fraction method has significant speed and accuracy advantages over the direct integration scheme and that this is why we have used it in the main text, since the cut does not alter the qualitative behaviour of the effect.

\subsubsection{Relative robustness of the RP spectrum}
A remarkable feature of the RP spectrum presented in Fig.~\ref{fig:RP_spectrum} is the fact that the ``original'' branch remains stable for the low overtones. 
Indeed, for the perturbation and the parameters considered, the QNMs spectrum, including the fundamental mode, is drastically modified. Here the fundamental RP and the first few overtones remain stable.
To quantify this, we introduce the absolute variation of the RP spectrum $\Delta \lambda_n = |\lambda_n^\epsilon - \lambda_n^0|$, where $\lambda_n^\epsilon$ is the perturbed RP and $\lambda_n^0$ the unperturbed one. 
Similarly, we introduce for the fundamental QNM frequency $\Delta \omega = |\omega_\epsilon - \omega_0|$, with $\omega_{\epsilon,0}$ referring to the fundamental QNM frequency.

\begin{figure}
    \centering
    \includegraphics[scale=0.65]{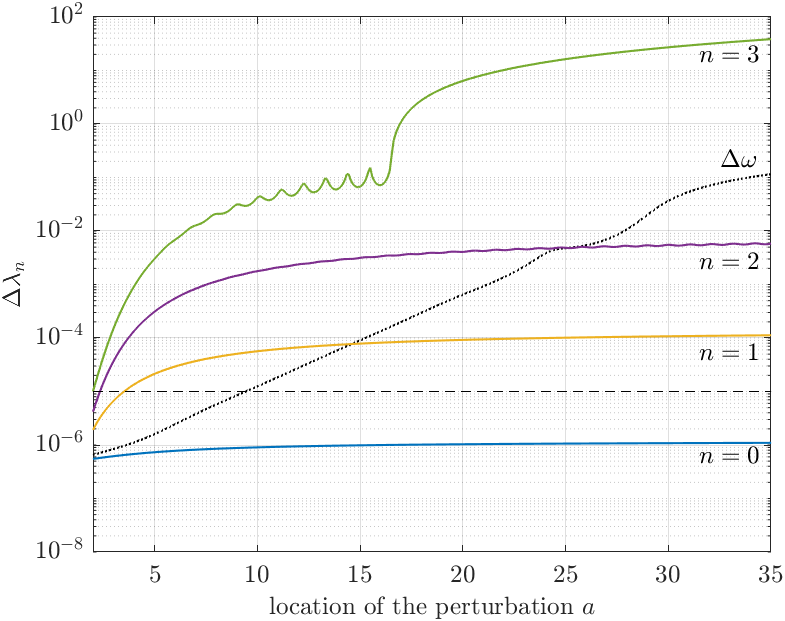}
    \caption{Superior robustness of the RP spectrum in comparison with the QNM frequency. The fundamental $n=0$ RP remains stable while the higher overtones are destabilised. The horizontal dashed line indicates the value of $\epsilon$ corresponding to the perturbation considered here.}
    \label{fig:robustness}
\end{figure}

Fig.~\ref{fig:robustness} depicts the evolution of $\Delta \lambda_n$ and $\Delta \omega$ as we vary the location of the perturbation. We can see that the fundamental RP ($n=0$) remains stable with a variation below $\epsilon$. The $n=1$ and $n=2$ RPs deviates from the unperturbed value by more than $\epsilon$ but the deviation seems to level off as $a$ increases. 
The $n=3$ RP, initially follows the trend of the lower overtones but gets destabilised after $a$ crosses a threshold. 
As a comparison we plot in dotted black, the variation of the QNM fundamental frequency which shows clear signs of destabilisation.
In contrast with the QNM spectrum, the fundamental Regge pole shows no sign of developing an instability for perturbations up to $a=35$. We can see that $|\Delta \lambda_0|$ actually reaches a constant and we therefore conjecture that the fundamental RP remains stable for all values of $a$.

\end{document}